\documentclass[review]{elsarticle}
\usepackage{booktabs}
\usepackage{array}
\usepackage{pdflscape}
\usepackage{amsmath}
\usepackage{amsfonts}
\usepackage{hyperref}
\usepackage{graphicx}
\graphicspath{{figures/}}
\usepackage{subfigure}
\usepackage{cite}
\usepackage{epstopdf}
\usepackage{multirow}
\usepackage{rotating}
\usepackage[mathlines]{lineno}
\usepackage{color}
\usepackage{lineno,hyperref}
\modulolinenumbers[5]

\journal{Journal of \LaTeX\ Templates}

\begin{document}

\begin{frontmatter}

\title{Hepatitis C Virus Epidemic Control Using a Nonlinear Adaptive Strategy}

	\author[a]{Javad Khodaei-Mehr}
	\author[b]{Samaneh Tangestanizadeh}
	\author[a]{Mojtaba Sharifi}
		\cortext[corr-author]{Corresponding author}
		\ead{Sharifi3@ualberta.ca}
	\author[b]{Ramin Vatankhah}
	\author[b]{Mohammad Eghtesad}

	\address[a]{School of Electrical and Computer Engineering, University of Alberta, Edmonton, Canada}
	\address[b]{School of Mechanical Engineering, Shiraz University, Shiraz, Iran}

\begin{abstract}
Hepatitis C is a viral infection that appears as a result of the Hepatitis C Virus (HCV), and it has been recognized as the main reason for liver diseases. HCV incidence is growing as an important issue in the epidemiology of infectious diseases. In the present study, a mathematical model is employed for simulating the dynamics of HCV outbreak in a population. The total population is divided into five compartments, including unaware and aware susceptible, acutely and chronically infected, and treated classes. Then, a Lyapunov-based nonlinear adaptive method is proposed for the first time to control the HCV epidemic considering modelling uncertainties. A positive definite Lyapunov candidate function is suggested, and adaptation and control laws are attained based on that. The main goal of the proposed control strategy is to decrease the population of unaware susceptible and chronically infected compartments by pursuing appropriate treatment scenarios. As a consequence of this decrease in the mentioned compartments, the population of aware susceptible individuals increases and the population of acutely infected and treated humans decreases. The Lyapunov stability theorem and Barbalat's lemma are employed in order to prove the tracking convergence to desired population reduction scenarios. Based on the acquired numerical results, the proposed nonlinear adaptive controller can achieve the above-mentioned objective by adjusting the inputs (rates of informing the susceptible people and treatment of chronically infected ones) and estimating uncertain parameter values based on the designed control and adaptation laws, respectively. Moreover, the proposed strategy is designed to be robust in the presence of different levels of parametric uncertainties. 
\end{abstract}


\end{frontmatter}

\section{Introduction}
\label{sec1}

The Hepatitis C Virus (HCV) is a blood-borne virus identified as the main cause of the liver disease [\citenum{JJJ,Prati,Wasley}]. Globally, about three percent of the world population (170 million) are dealing with the HCV and 71 million people have chronic hepatitis C infection [\citenum{JJJ,Zhang,ZhangZhou,Chen}].  Several studies showed that the chronic stage of HCV will develop the cirrhosis and liver cancer in the case of no treatment and approximately 339 thousands people die every year due to these diseases[\citenum{JJJ,Bisceglie}].
Despite previously mentioned statistics which makes HCV infection as one of the important health threats, this disease received little attention especially in the regions with higher rate of infectiousness [\citenum{Zhang}].\\

Although fatigue and jaundice were mentioned as symptoms of the HCV, but this disease often has no considerable symptom, even in the advanced stages. This is the reason that the HCV outbreak is called "the silent epidemic" [\citenum{Zhang,YuanYang}]. 
Several different ways were reported for HCV prevalence, which include sharing injection equipments, unsafe sexual contacts, inadequately sterilization of syringes and needles especially for health-care personnel and transfusion of polluted blood [\citenum{JJJ,Klevens}]. Even though these are the main causes of HCV epidemic, but some other reasons may also be critical in some societies based on special conditions. For instance in developed countries, since there is precise control on the blood transfusion, the importance of injecting drug use in transmission of the disease has increased compared to the transfusion of polluted blood and its products [\citenum{Prati,Klevens}].\\

Natural cure at the chronic stage of HCV is not common, but it can happen for about 10-15\% of patients that the RNA of HCV is indistinguishable in their serum [\citenum{ZhangZhou,Chen}]. For the rest of patients (80-85 \%) that the HCV could not be healed by their immune system response, some drug therapy regimes should be employed. Hepatitis C drugs have recently had some developments. Available safe, highly effective and endurable combinations of oral antivirals that act directly, have currently developed for this disease [\citenum{Zhang,Banerjee}]. Although vaccination is the vital way of controlling different viral diseases, but unfortunately there is no vaccine for the HCV yet [\citenum{ZhangZhou}]. Therefore, preventing this disease has an  important role in stopping the extension of the its epidemic.\\

In the present work, a nonlinear adaptive method is developed for treatment and control of the HCV epidemic. For this purpose, the recently published nonlinear HCV epidemiological model in [\citenum{Zhang}] is employed and different parametric uncertainties are taken into account, despite the previous optimal strategies [\citenum{Zhang}]. The main goal of the proposed control scheme is the population decrease in the unaware susceptible and chronically infected compartments in the existence of parametric uncertainties. Accordingly, two control inputs (efforts to inform susceptible individuals and treatment rate) are employed to track descending desired populations of the previously mentioned compartments. The asymptotic stability and tracking convergence of the closed-loop system having uncertainties are proven using the Lyapunov stability theorem and the Barbalat's lemma. Innovations of this research are as follows:
\begin{itemize}
	\item For the first time a nonlinear adaptive method is developed to control the HCV epidemic by defining a novel Lyapunov function candidate that provides the tracking convergence proof. 
	\item Due to the lack of accurate information about HCV model parameters in each society, parametric uncertainty is taken into account in this research for the first time, and the defined control objectives are achieved in the presence of these inaccuracies.
	\item In all of the previous studies that have been conducted on the control of the HCV outbreak, populations of some undesired compartments at the end of investigation period were considered as the criterion for designing control inputs [\citenum{Oare,Okosun,Zhang,khodaei,khodaei2}]. However, in the present work for the first time, the populations of two unawared susceptible and chronically infected classes during the entire treatment period are considered as the criterion and control inputs are designed in a way to track desired values instead of focusing on their final populations at the end of process.
\end{itemize} 
The rest of this chapter is organized as follows. In Sec. 2, related research work will be explained. Description of the dynamic model and the proposed control scheme will be presented in Sec. 3 and Sec. 4, respectively. Results of simulations will be depicted and discussed in Sec. 5, and the concluding remarks will be mentioned in Sec. 6.

\section{Related research work}
\label{Related}
Previous related studies are presented in this section and divided into three parts, including mathematical modeling, and Optimal control for HCV and adaptive control strategies for different biological systems.\\

Several analytical analyses were conducted on the dynamic modeling of the HCV epidemic which are presented here. Martcheva and Chavez [\citenum{Martcheva}] presented a simple mathematical model with three compartmental variables including susceptible, acutely infected and chronically infected. They considered different epidemiological observations in the model. Yuan and Yang [\citenum{YuanYang}] added the exposed class to the previous model [\citenum{Martcheva}]. They considered that the susceptible individuals transmit to the exposed compartment in the case of having contact with the infected compartments. Zhang and Zhou [\citenum{ZhangZhou}] added a new term in the model of Yuan and Yang [\citenum{YuanYang}], which denotes the death rate due to the HCV. Hu and Sun [\citenum{Hu}] proposed another epidemiological model for the HCV with four classes in which the recovered compartment was taken into account for the first time. Naturally the recovered people transmit to this class from the acutely infected and chronically infected compartments and become immune against this. Ainea et al. [\citenum{Ainea}] extended the previous model [\citenum{Hu}] by adding the exposed class. Both of these models [\citenum{Hu}] and [\citenum{Ainea}] considered the HCV disease-induced death rate for both acutely-infected and chronically infected classes. Shen et al. [\citenum{Shen}] proposed a dynamical model with six classes including susceptible, exposed, acutely infected, chronically infected, treated and recovered populations. They propounded treatment influence for the first time and classified treated people in the distinct class. Shi and Cui [\citenum{Shi}] improved the model in [\citenum{Shen}] and divided the treated class into two different classes by defining the treatment for chronical infection and awareal reinfection. \\

Some researches have been conducted for optimal control of the HCV outbreak. Okosun [\citenum{Oare}] employed a SITV (susceptible, acutely infected, treated and chronically infected) model for the HCV that was an extended form of the dynamics presented in [\citenum{YuanYang}]. This model [\citenum{Oare}] included the treatment compartment and considered movement for susceptible, treated, and acutely and chronically infected people among their compartments. Some time dependent optimal control strategies are proposed, in order to control the HCV disease. Cost function is calculated for these strategies in order to evaluate effectiveness of the control methods and select the most efficient one. 
Okosun and Makinde [\citenum{Okosun}] employed a SEITV (susceptible, exposed, acutely infected, treated and chronically infected) dynamical model for the HCV outbreak considering the screening rate and drug efficacy as control inputs for acutely and chronically infected populations and used the Pontryagin's Principle to solve the optimal control problem.
Another epidemiological model was investigated in [\citenum{Zhang}] for the HCV outbreak in which the susceptible class was divided into aware and unaware classes. Moreover, they considered two control inputs including screening and treatment rates for the HCV epidemic model which were determined by an optimal control law. In [\citenum{Zhang}], the dynamics was formulated with the susceptibility reduction due to the publicity and the treatment process to identify the feasible effect of public concerns and treatment on the HCV.
An optimal neuro-fuzzy strategy was also introduced in  [\citenum{khodaei}] in order to control the HCV epidemic. They [\citenum{khodaei}] employed the mathematical model proposed in [\citenum{Okosun}] as a deterministic model and utilized the genetic algorithm to obtain optimal control inputs. \\

As described, all of previous studies on the control of HCV epidemic were conducted on the optimal strategies. On the other hand, some other research works were performed on the adaptive control of different diseases as presented here. 
Moradi et al. [\citenum{Moradi}] suggested a Lyapunov-based adaptive method to control three different  hypothetical models of the cancer chemotherapy inside the human body and compared results among these models. In the next step of this research [\citenum{Sharifi2}], a composite adaptive strategy has been developed for online identification of cancer parameters during the chemotherapy process. 
Boiroux et al. [\citenum{Boiroux}] employed a model predictive controller for the type 1 diabetes' model and used an adaptive controller to balance the blood glucose. They determined the model parameters based on clinical information of past patients. Aghajanzadeh et al. [\citenum{Aghajanzadeh}] suggested an adaptive control strategy for hepatitis B virus infection inside the human body by antiviral drugs. They considered model parameters uncertainties on model parameters and employed adaptive controller to control the dynamic despite uncertainties of the system. Sharifi and Moradi[\citenum{Sharifi}] designed a robust scheme with adaptive gains to control the influenza epidemic, considering its dynamic model's uncertainties. Padmanabhan et al. [\citenum{Padmanabhan}] proposed an optimal adaptive method to control the sedative drug in anesthesia administration. They employed an integral reinforcement learning method in order to overcome the uncertainty of parameter values.\\

\section{Dynamic model of hepatitis C virus epidemic}
\label{sec2}

Mathematical modeling is an useful way of analysis for epidemiology of a disease. These models have two important capabilities: 1. finding out mechanistic understanding of the disease, and 2. exploring potential outcomes of the epidemic under different conditions [\citenum{Probert}]. For assessment of the proposed method for the HCV prevalence control in a population, a nonlinear compartmental model is used with five different classes including  unaware susceptible (\textit{$S_u$}), aware susceptible (\textit{$S_a$}), acutely infected (\textit{I}), chronically infected (\textit{C}) and the treated (\textit{T}) humans [\citenum{Zhang}]. The susceptible compartment is divided into two classes, including aware and unaware people. Note that aware people have information about the HCV transmission ways and preventing methods despite unaware population. Since there is no available vaccine for the HCV, informing people about preventing methods is a very important way to reduce the risk of infection for susceptible people [\citenum{JJJ}]. Therefore, the unaware susceptible individuals (\textit{$S_u$}) will be infected in contact with the infected population (\textit{I, C} and \textit{T}) with a higher rate in comparison with the aware susceptible individuals (\textit{$S_a$}) [\citenum{Zhang}]. Thus, the transmission rate for unaware susceptible humans (\textit{$S_u$}) should be considered larger than this rate for aware susceptible humans (\textit{$S_a$}) in the dynamic model [\citenum{Zhang}]. The nonlinear mathematical model of HCV epidemic is as follows: 
\begin{align}
&\dot{S_u}=b-\lambda_{S_u}\frac{S_u}{N}-(\mu+u_1(t))S_u+(1-q)\gamma I \nonumber \\
&\dot{S_a}=u_1(t) S_u-\lambda_{S_a}\frac{S_a}{N}-\mu S_a+(1-p)\xi T  \nonumber  \\
&\dot{I}=\lambda_{S_u}\frac{S_u}{N}+\lambda_{S_a}\dfrac{S_a}{N}-(\mu + \gamma)I \label{DynEq}  \\ 
&\dot{C}=q \gamma I-(\mu + u_2(t) + \theta)C + p \xi T      \nonumber \\
&\dot{T}=u_2(t) C- (\mu + \xi)T   \nonumber
\end{align}
where $\lambda_{S_u}=\beta (I+K_1 C + K_2 T)$ and $\lambda_{S_a}=\alpha \lambda_{S_u}$. $u_1$ and $u_2$ are control inputs and defined as effort rate to inform unaware susceptible individuals and treatment rate for chronically infected class, respectively. \textit{N} denotes the total population and will be calculated as:
\begin{align}
N=S_u+S_a+I+C+T
\end{align}

The population of unaware susceptible (\textit{$S_u$}) increases with the rate of \textit{b}. Unaware and aware susceptible individuals are also infected in contact with acutely and chronically infected and treated individuals at the rates of $\lambda_{S_u}$ and $\lambda_{S_a}$, respectively. Infectiousness rate for acutely infected people is higher than chronically infected individuals, and the treated people have the lowest rate; thus, it is assumed that $K_1>K_2$ [\citenum{Zhang,ZhangZhou}]. The total population (\textit{N}) decreases with two different rates $\mu$ and $\theta$, where $\mu$ denotes the rate of natural death that decreases populations of different compartments. However, $\theta$ is the rate of HCV induced death and decreases the population of the chronically infected compartment (\textit{C}). \\

\begin{figure*}
	\begin{center}
	\includegraphics[width=0.75\textwidth]{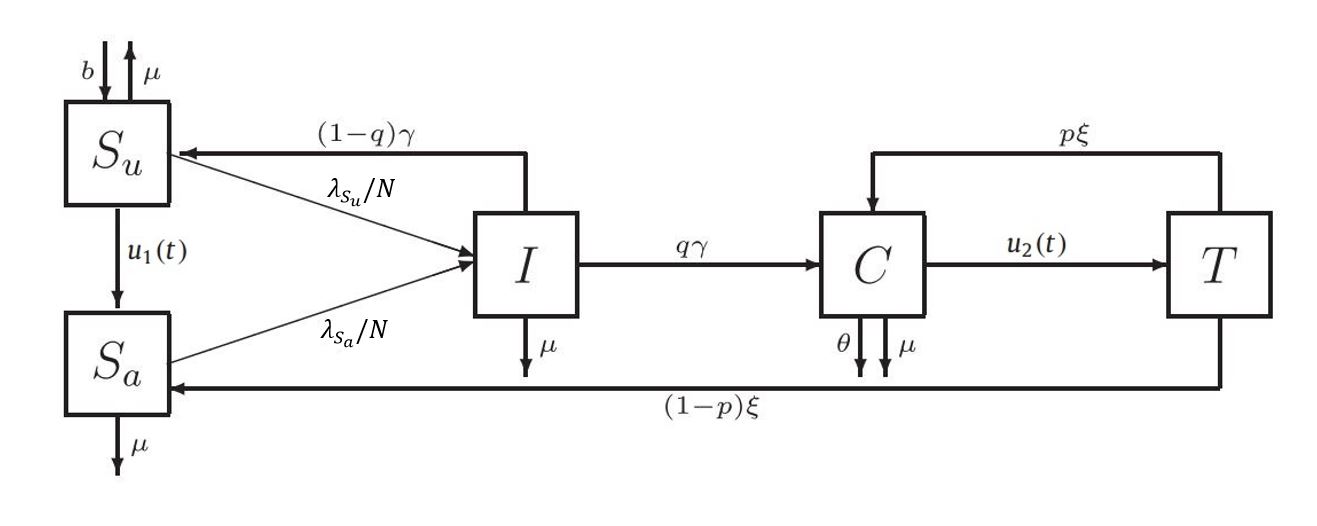}
	\caption{Schematic diagram of transition among different classes of HCV epidemic} 
	\label{SchModel}
	\end{center}
\end{figure*}

During the acute stage (\textit{I}), the HCV could have different behaviors for each patient based on his/her immune system response. For 15 to 25\% of cases in this stage, the RNA of HCV becomes indistinguishable in their blood serum and the ALT level returns to the normal range. This observation is defined by the term $(1-q)\gamma I$ in the proposed HCV dynamics [\citenum{Zhang,Chen}]. Approximately, the immune system in 75-85\% of the patients could not remove the hepatitis C virus in the acute stage and their disease becomes advanced to the chronic stage. Note that if the HCV RNA reamins in the patient's blood for at least six months after onset of acute infection, the chronic level of the disease will appear which is defined by the term $q\gamma I$ in Eq. (\ref{DynEq}) [\citenum{ZhangZhou,Chen}]. Finally, the defeat in the treatment process is defined by the term \textit{p}. The treated population decreases by the rate of $\xi T$ and join the chronic class by the rate of $p \xi T$ in the case of treatment defeat and the rest of this population $(1-p) \xi T$ will join aware susceptible class if the treatment be successful. The schematic diagaram of the proposed nonlinear dynamics of the HCV epidemic is depicted in the Fig. \ref{SchModel} and descriptions of the parameters are presented in Table \ref{ParDef} [\citenum{Zhang}].\\

\begin{table}
	\caption{Parameters of the mathematical model of the HCV (\ref{DynEq}) [\citenum{Zhang}]} \label{ParDef}
	\begin{tabular}{ll}
		\hline
		Parameter & Description \\
		\hline
		b         & Rate of birth \\
		$\mu$     & Rate of death \\
		$\beta$   & Transmission coefficient \\
		$K_1$     & Chronic stage infectiousness relative to acute stage \\
		$K_2$     & Treated individuals' infectiousness relative to acute ones \\
		$\alpha$  & Rate of being infected for aware people relative to unaware ones \\
		$\gamma$  & Leaving rate of acutely infected class \\
		$q$       & Progressing proportion from acute stage to chronic one \\
		$\xi$     & Transferring rate from treated class to other ones  \\
		$p$       & Moving back proportion from treated class to chronic one \\
		$\theta$  & HCV induced death rate \\
		\hline
	\end{tabular}
\end{table}

\section{Nonlinear adaptive controller formulation for epidemiology of HCV}
\label{sec3}

In the present section, a new nonlinear adaptive controller is formulated for uncertain hepatitis C virus epidemic. The main purpose of the control method is to minimize the populations of unaware susceptible (\textit{$S_u$}) and chronically infected (\textit{C}) classes. Two control inputs $u_1(t)$ and $u_2(t)$ are considered in order to reach this objective. $u_1(t)$ denotes the effort rate to inform the susceptible individuals from the HCV by media publicity, educational campaigns, public service advertising and so on, and $u_2(t)$ is employed to reflect the rate of treatment on chronically infected individuals [\citenum{Zhang}]. \\

\begin{figure}
\begin{center}
\includegraphics[scale=0.45]{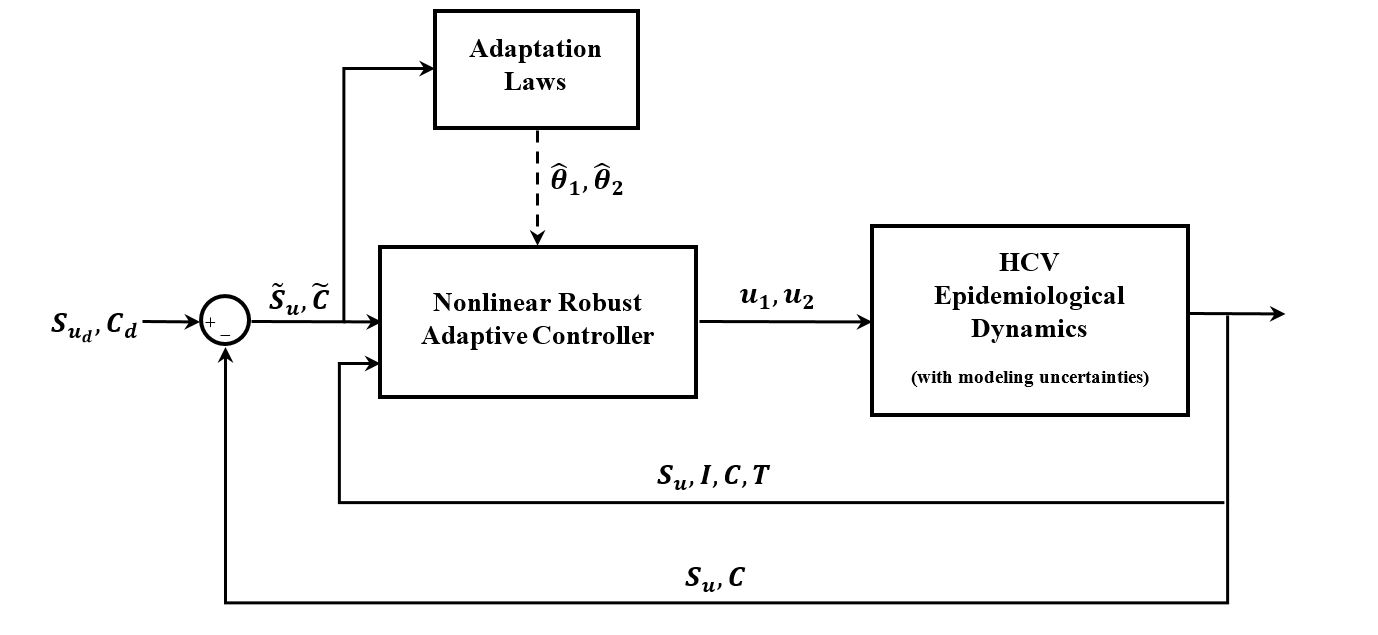}
\caption{Conceptual diagram of nonlinear adaptive method developed to control the HCV epidemic in the existence of uncertainties on parameters of the model} \label{SchControl}
\end{center}
\end{figure}

Using the above-mentioned control inputs, the populations of unaware susceptible (\textit{$S_u$}) and chronically infected (\textit{C}) classes will decrease by tracking some desired values. Moreover, due to decrease of the mentioned components, the number of aware susceptible (\textit{$S_a$}) and treated (\textit{T}) individuals will increase and decrease, respectively. The Lyapunov theorem is employed to prove stability of the closed-loop system. In addition, some adaptation laws are defined in order to update estimated parameters of the system to guarantee the stability and robustness of the system against uncertainties of the dynamic model. A conceptual diagram of the proposed nonlinear feedback controller with adaptive scheme is illustrated in Fig. \ref{SchControl}. \\

\subsection{Nonlinear adaptive control laws}
\label{subsec3.1}

Control inputs ($u_1(t)$, $u_2(t)$) could be calculated using dynamics of the unaware susceptible and chronically infected compartments from Eq. (\ref{DynEq}) as:
\begin{align}
u_1&=-\frac{\dot{S_u}}{S_u}+\frac{b}{S_u}-\frac{\beta}{N} (I+K_1C+K_2T)-\mu+(1-q)\gamma \frac{I}{S_u} \label{u1} \\
u_2&=-\frac{\dot{C}}{C}+q \gamma \frac{I}{C}-(\mu + \theta) +  p\xi \frac{T}{C} \label{u2}   
\end{align}
\\
\textbf{Property.} The right-hand sides of Eqs. (\ref{u1}) and (\ref{u2}) can be linearly parameterized in terms of their available parameters. $\phi_1$ and $\phi_2$ are considered to be the arbitrary variables instead of $\dot{S_u}$ and $\dot{C}$. Now the equations could be rewritten as:

\begin{align}
-\frac{\dot{S_u}}{S_u}+\frac{b}{S_u}-\frac{\beta}{N} (I+K_1C+K_2T)-\mu+(1-q)\gamma \frac{I}{S_u} \label{u1par}&= -\frac{\phi_1}{S_u}+Y_1 \theta_1 \\
-\frac{\dot{C}}{C}+q \gamma \frac{I}{C}-(\mu + \theta) +  p\xi \frac{T}{C}= -\frac{\phi_2}{C}+Y_2 \theta_2 \label{u2par}   
\end{align}
where $Y_1$ and $Y_2$ are the regressor matrices contain known functions of HCV epidemic variables. $\theta_1$ and $\theta_2$ are parameter vectors which contain unknown parameters of the dynamic. Eqs. (\ref{RegPar1}) and (\ref{RegPar2}). Accordingly, these matrices and vectors are defined as

\begin{align}
Y_1=
&\begin{bmatrix}
\dfrac{1}{S_u} & -\dfrac{I}{N} & -\dfrac{C}{N} & -\dfrac{T}{N} & \dfrac{I}{S_u} & -1
\end{bmatrix}; 
\quad \theta_1= 
\begin{bmatrix}
b & \beta & \beta K_1 & \beta K_2 & (1-q)\gamma & \mu
\end{bmatrix}^T \label{RegPar1}\\
Y_2=
&\begin{bmatrix}
\dfrac{I}{C} & \dfrac{T}{C} & -1
\end{bmatrix};
\quad \theta_2= 
\begin{bmatrix}
q \gamma & p \xi & (\mu + \theta) 
\end{bmatrix}^T \label{RegPar2}
\end{align} 

This regressor presentation is used to summarize the equations and define the adaptation and control laws. In order to design nonlinear control laws, two new variables $\phi_1$ and $\phi_2$ are defined as follows:
\begin{align}
\phi_1&=\dot{S}_{u_d}-\lambda_1 \tilde{S}_u \\
\phi_2&=\dot{C}_d-\lambda_2 \tilde{C} 
\end{align}
where $\lambda_1$ and $\lambda_2$ are the controller gains and considered to be positive and constant. Now, the nonlinear adaptive control laws are defined as
\begin{align}
u_1&=-\frac{\dot{S_{u_d}}-\lambda_1 \tilde{S_u}}{S_u}+Y_1 \hat{\theta}_1 \label{ConLaw1}  \\
u_2&=-\frac{\dot{C_d}-\lambda_2 \tilde{C}}{C}+Y_2 \hat{\theta}_2 \label{ConLaw2}  
\end{align}
where $\hat{\theta}_1$ and $\hat{\theta}_2$ are the vectors of estimated parameters.\\

In the next section, taking advantages of the Lyapunov stability theorem, it will be proven that the control laws (\ref{ConLaw1}) and (\ref{ConLaw2}) together with some adaptation laws, guarantee the tracking convergence, stability and robustness for the treatment of HCV outbreak.

\subsection{Stability proof and adaptation laws}
The closed-loop dynamics of the system is achieved firstly by substituting the control laws (\ref{ConLaw1}) and (\ref{ConLaw2}) into the dynamics of HCV epidemic (\ref{DynEq}):
\begin{align}
\frac{\dot{\tilde{S}}_u+\lambda_1 \tilde{S}_u}{S_u}&=-Y_1 \tilde{\theta}_1 \label{ClsLp1}\\
\frac{\dot{\tilde{C}}+\lambda_2 \tilde{C}}{C}&=-Y_2 \tilde{\theta}_2 \label{ClsLp2}
\end{align}
where $\tilde{\theta}_i$ (for \textit{i}=1, 2) is defined as $\hat{\theta}_i-\theta_i$.\\

The adaptation laws are designed to update parameters' estimation to keep the system's robustness against uncertainties, as
\begin{align}
\dot{\hat{\theta}}_1^T&=S_u \tilde{S}_u Y_1 \Gamma_1 \label{AdpLaw1}\\
\dot{\hat{\theta}}_2^T&=C \tilde{C} Y_2 \Gamma_2 \label{AdpLaw2}
\end{align}
where $\Gamma_1$ and $\Gamma_2$ are the adaptation gain matrices and considered to be positive definite.\\

Now, employing the Lyapunov stability theorem [\citenum{Slotine}] and based on the previously derived close-loop dynamics (\ref{ClsLp1})-(\ref{ClsLp2}) and the designed adaptation laws (\ref{AdpLaw1})-(\ref{AdpLaw2}), the tracking convergence, stability and robustness for aware susceptible and chronically infected classes will be proven. With this aim, a positive definite Lyapunov-candidate-function is selected as
\begin{align}
V=\frac{1}{2}[\tilde{S}_u^2+\tilde{C}^2+\tilde{\theta}_1^T \Gamma_1^{-1}\tilde{\theta}_1 + \tilde{\theta}_2^T \Gamma_2^{-1}\tilde{\theta}_2] \label{Lyapunov}
\end{align}

The Lyapunov function's time derivative is determined:
\begin{align}
\dot{V}=\tilde{S}_u\dot{\tilde{S}}_u +\tilde{C}\dot{\tilde{C}} + \dot{\hat{\theta}}_1^T \Gamma_1^{-1}\tilde{\theta}_1 +  \dot{\hat{\theta}}_2^T \Gamma_2^{-1}\tilde{\theta}_2  \label{Vdot}
\end{align}

It should be mentioned that $\dot{\tilde{\theta}}=\dot{\hat{\theta}}$ because $\theta$ is constant ($\dot{\theta}$ is zero). By substituting the adaptation laws (\ref{AdpLaw1}) and (\ref{AdpLaw2}) into (\ref{Vdot}), the time derivative of V is simplified to:
\begin{align}
\dot{V}=-\lambda_1 \tilde{S}_u^2 -\lambda_2 \tilde{C}^2  \label{Vdot2}
\end{align}

As mentioned in the previous descriptions, $\lambda_1$ and $\lambda_2$ are considered to be positive; thus, the Lyapunov function's time derivative is negative-semi-definite. Thus, based on the Barbalat's Lemma (described in the Appendix A) and the Lyapunov stability theorem [\citenum{Slotine}], it is proven that $\tilde{S}_u$ and $\tilde{C}$  converge to the zero. In other words, employing the suggested adaptive feedback control strategy ensures the tracking convergence and stability ($\tilde{S}_u \rightarrow 0$ and $\tilde{C} \rightarrow 0$ as $t\rightarrow \infty$) in the presence of uncertainties. Thus, the numbers of unaware susceptible (\textit{$S_u$}) and chronically infected (\textit{C}) converge to the desired values ($S_u\rightarrow S_{u_{d}}$ and $C\rightarrow C_{d}$). \\

\section{Results and discussion}
For effectiveness evaluation of the proposed method, some simulations are presented in this section. Note that computer simulations have proven to be useful for evaluating the spread behavior of infectious diseases [\citenum{Orbann}]. In the present study, the simulation process is performed in the Simulink-Matlab environment. The parameters' values of the HCV epidemic model (\ref{DynEq}) are listed in Table \ref{ParVal}. 

\begin{table}
\begin{center}
\caption{Values of the HCV parameters in its mathematical model (\ref{DynEq}) [\citenum{Zhang}]} \label{ParVal}
 \begin{tabular}{lc}
 \hline
 Parameter & Value \\
 \hline
 b         & 0.012 \\
 $\mu$     & 0.006 \\
 $\beta$   & 0.15 \\
 $K_1$     & 0.5 \\
 $K_2$     & 0.2 \\
 $\alpha$  & 0.1 \\
 $\gamma$  & 4 \\
 $q$       & 0.2 \\
 $\xi$     & 0.8 \\
 $p$       & 0.5 \\
 $\theta$  & 0.001 \\
 \hline
 \end{tabular}
 \end{center}
 \end{table}
 
 A small society with total population of 1310 people at the beginning of investigation is used. The following desired scenarios are considered for reduction of unaware susceptible individuals (\textit{$S_{u_d}$}) and treatment of chronically infected people (\textit{$C_d$}).
 
\begin{align}
S_{u_d}&=(S_{u_0}-S_{u_f})exp(-a_1t)+ S_{u_f} \label{Desired1} \\
C_d&=(C_0-C_f)exp(-a_2t)+ C_f  \label{Desired2}
\end{align}
where \textit{$a_1$} and \textit{$a_2$} are the desired population reduction rates. \textit{$S_{u_0}$} and \textit{$S_{u_f}$} are the initial and final (steady state) populations of unaware susceptible class, respectively.  \textit{$C_0$} and \textit{$C_f$} are the initial and final (steady state) populations of chronically infected compartment, respectively.\\

\begin{figure}
\begin{center}
\includegraphics[scale=0.75]{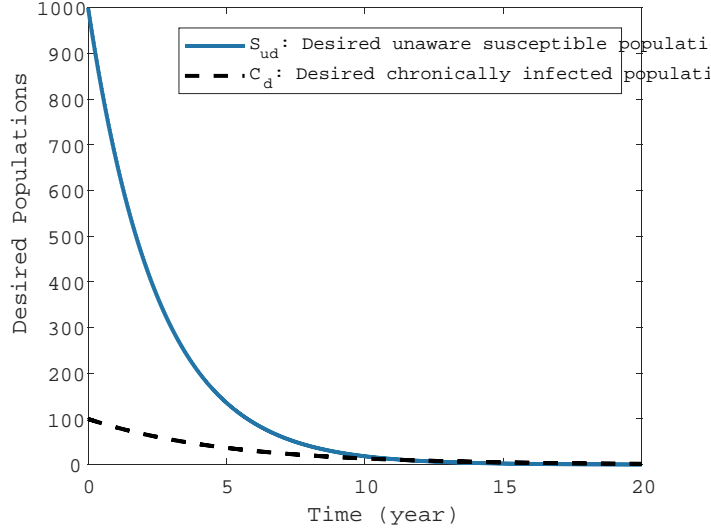}
\caption{Desired scenarios for reduction of unaware susceptible and chronically infected compartments in the HCV epidemic} \label{Desired}
\end{center}
\end{figure}

The presented reduction and treatment scenarios (\ref{Desired1}) and (\ref{Desired2}) are employed in these simulations as the desired decreasing behavior of the HCV epidemic control. However, other continuously decreasing fashion can be used as desired scenarios without loss of generality. Values of parameters in the desired HCV population reduction scenarios (~\ref{Desired1}) and (\ref{Desired2}) are listed in Table \ref{DesVal}. These scenarios for unaware susceptible and chronically infected compartments are shown in Fig. \ref{Desired}.\\

\begin{table}
\begin{center}
\caption{Values of parameters in the desired HCV population reduction scenarios (\ref{Desired1}) and (\ref{Desired2})} \label{DesVal}
 \begin{tabular}{lc}
 \hline
 Parameter     & Value \\
 \hline
 $S_{u_0}$     & 1000 \\
 $C_0$         & 100 \\
 $S_{u_f}$     & 0 \\
 $C_f$         & 0 \\
 $a_1$         & 0.4 \\
 $a_2$         & 0.2 \\
 \hline
 \end{tabular}
 \end{center}
 \end{table}
 
\begin{figure}
\begin{center}
     {\includegraphics[scale=0.75]{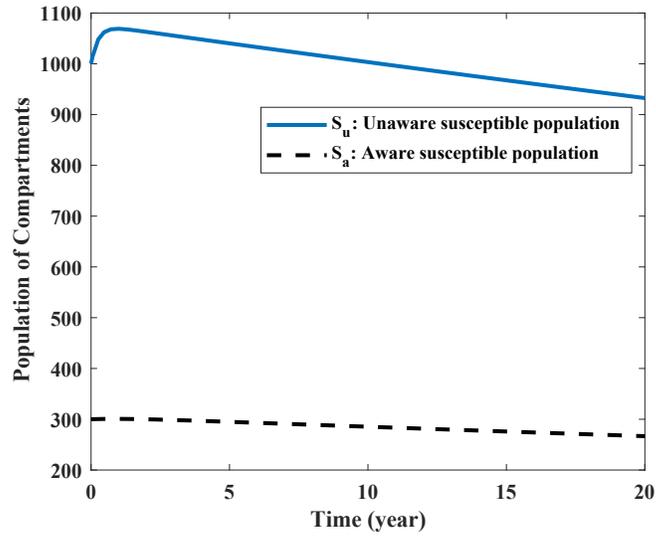}}\label{No control1}\\
     (a)\\
     { \includegraphics[scale=0.75]{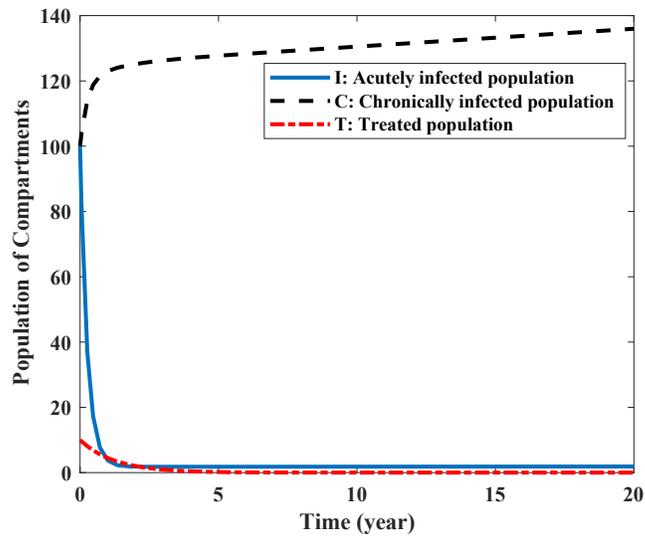}}\label{No control2}\\
     (b)
\caption{Populations of (a) unaware and aware susceptible, and (b) acutely infected, chronically infected and treated classes in the absence of control inputs} \label{No control}
\end{center}
\end{figure}

In the absence of control inputs, the HCV infection will extend in the society based on Eq. \ref{DynEq}. Accordingly, the treated population will decrease and reach zero exponentially due to the lack of treatment process. In that case (no control input), unaware and aware susceptible individuals will get the hepatitis C virus in contact with the infected people in \textit{I} and \textit{C} compartments and will join the acutely infected class (\textit{I}). Since there is no treatment for acutely infected individuals (as seen in Eq. (\ref{DynEq})), the HCV disease will progress and reach the chronic stage. Thus, the population of the chronically infected compartment (\textit{C}) will increase and the populations of all other compartments will decrease. Figure \ref{No control} depicts the above-mentioned points about the HCV outbreak in the case of no control input.\\

\begin{figure}
\begin{center}
     {\includegraphics[scale=0.75]{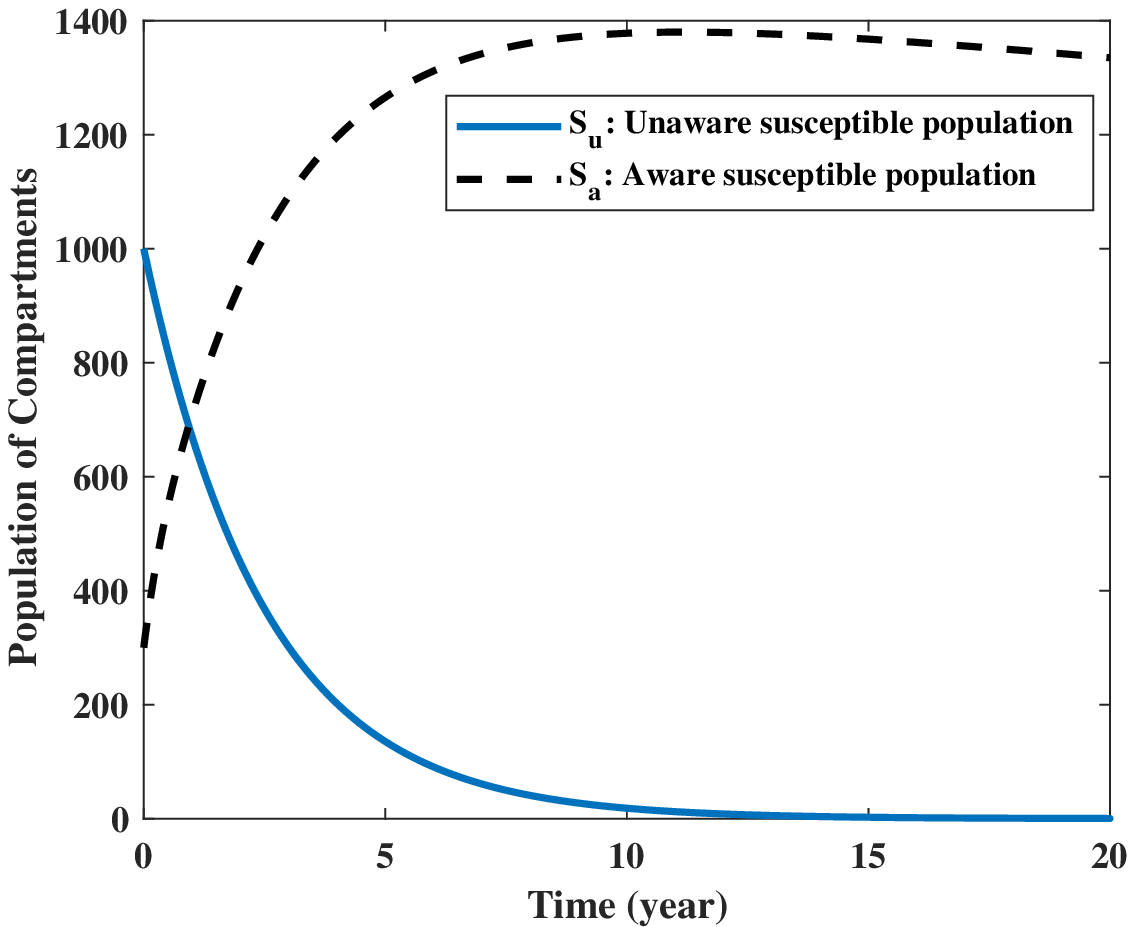}}\label{With control1}\\
     (a)\\
     {\includegraphics[scale=0.75]{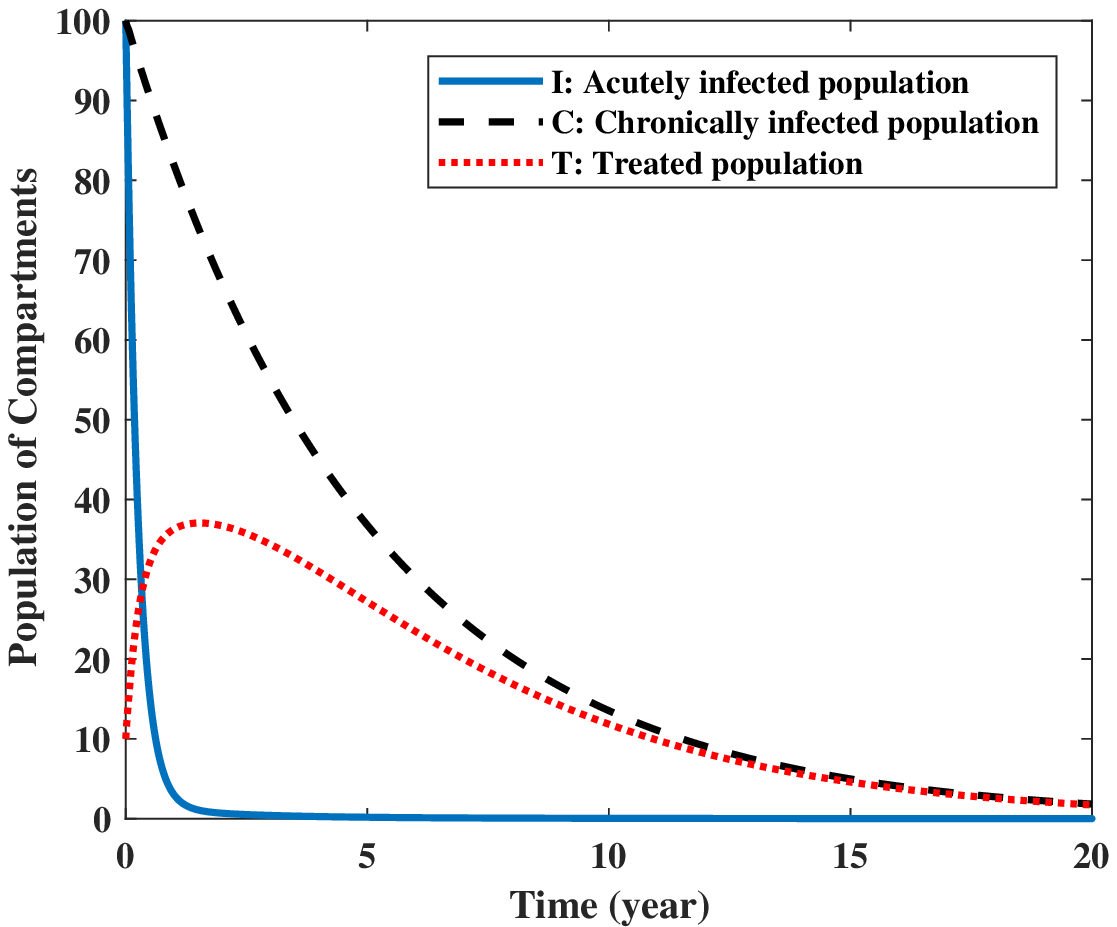}}\label{With control2} \\
     (b)

\caption{Populations of (a) unaware and aware susceptible, and (b) acutely infected, chronically infected and treated classes in the presence of control inputs ($u_1$ and $u_2$) based on the proposed laws (\ref{ConLaw1}) and (\ref{ConLaw2})} \label{With control}
\end{center}
\end{figure}

However, applying the proposed strategy based on the designed nonlinear control laws (\ref{ConLaw1}) and (\ref{ConLaw2}) with the obtained adaptation laws (\ref{AdpLaw1}) and (\ref{AdpLaw2}), the population changes in different compartments in the presence of 20\% parametric uncertainty are depicted in Fig. \ref{With control}.\\

\begin{figure}
\begin{center}
\includegraphics[scale=0.75]{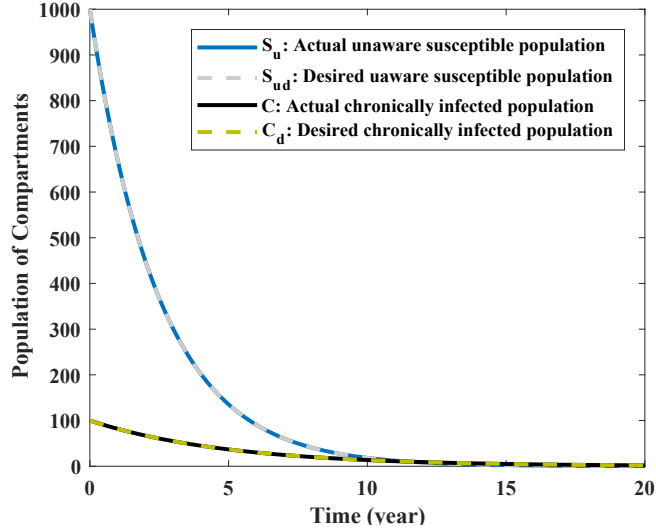}
\caption{Convergence of unaware susceptible and chronically infected populations ($S_u$ and \textit{C}) to their desired values ($S_{u_d}$ and $C_d$)} \label{Convergence}
\end{center}
\end{figure}

\begin{figure}
\begin{center}
\includegraphics[scale=0.75]{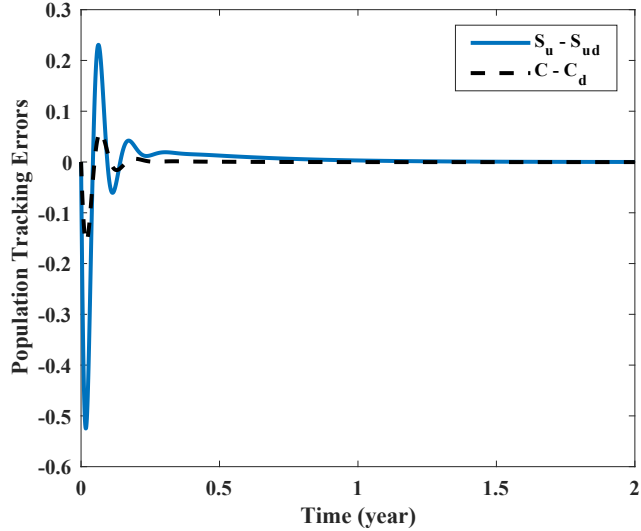}
\caption{Tracking errors between the desired and real values of unaware susceptible and acutely infected compartments} \label{Error}
\end{center}
\end{figure}

As seen, due to the employment of the first control input (\textit{$u_1$}), the unaware susceptible individuals (\textit{$S_u$}) reduce and join the aware susceptible class (\textit{$S_a$}). Since the aware susceptible people become less infectious than unaware ones because of the control input $u_1$, extension of the HCV infection decreases compared with the no-control-input case (shown in Fig. \ref{No control}). Moreover, using the treatment rate as the second control input (\textit{$u_2$}), the population of chronically infected compartment (\textit{C}) decreases (Fig. \ref{With control}) based on the described scenarios ($C_d$ in Fig. \ref{Desired}). Thus, the populations of unaware susceptible and chronically infected classes reduce and the population of aware susceptible increases in Fig. \ref{With control}, which are in accordance with the HCV dynamics (\ref{DynEq}). Although 20\% parametric uncertainty is taken to account for the nonlinear model, simulation results show that the proposed control strategy satisfied its objective which is convegence to desired population reduction and treatment scenarios ($S_u \rightarrow S_{u_d}$ and $C \rightarrow C_d$). Figure \ref{Convergence} depicts the desired and real populations of unaware susceptible and chronically infected classes,which implies the appropriate convergence performance using the nonlinear controller. The corresponding tracking errors are presented in Fig. \ref{Error}. \\
 
\begin{figure}
\begin{center}
\includegraphics[scale=0.75]{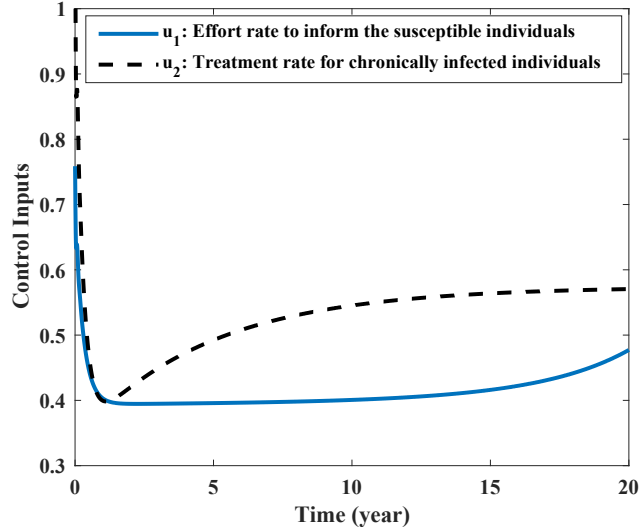}
\caption{Control inputs ($u_1$ and $u_2$) during the treatment period of HCV epidemic} \label{Control inputs}
\end{center}
\end{figure}

\begin{figure}
\begin{center}
	 {\includegraphics[scale=0.75]{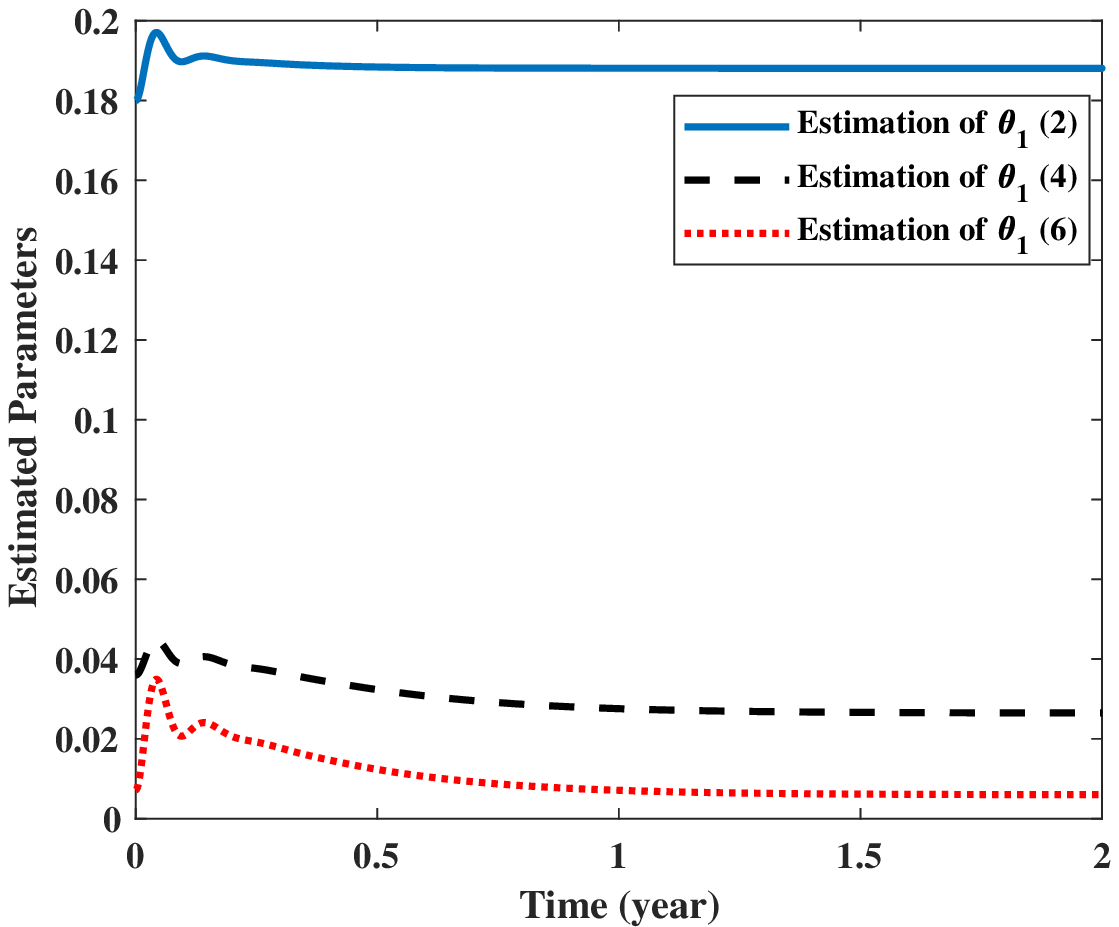}}\label{Par1}\\
	 (a)\\
     {\includegraphics[scale=0.75]{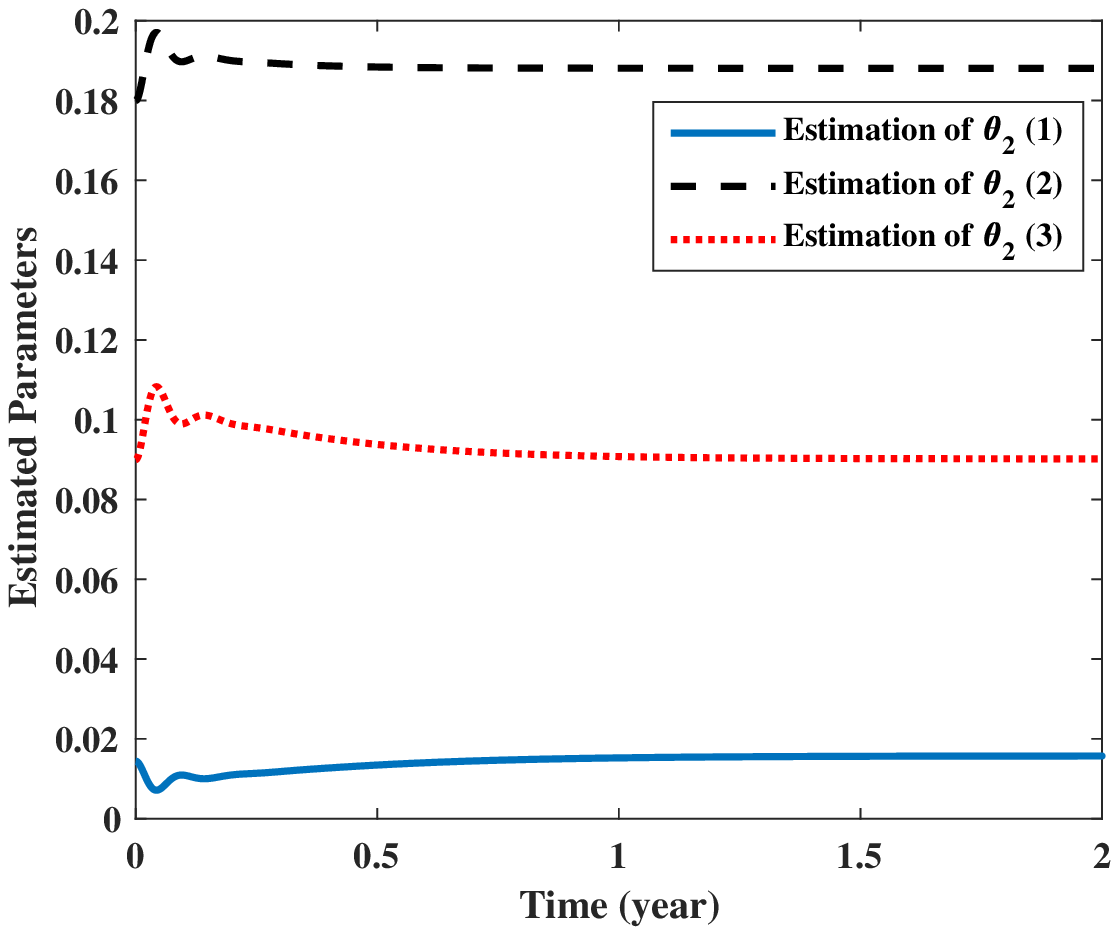}}\label{Par2} \\
     (b)

\caption{Estimation of parameters in (a) $\theta_1$ and (b) $\theta_2$ during the treatment period based on the adaptation laws (\ref{AdpLaw1}) and  (\ref{AdpLaw2}), respectively} \label{Par}
\end{center}
\end{figure}

 As described, two control inputs are adjusted according to the proposed nonlinear adaptive strategy in order to prevent the HCV outbreak. The first control input $u_1(t)$ denotes the effort rate to inform the susceptible individuals from the HCV and the second one $u_2(t)$ is the treatment rate for chronically infected individuals. These control inputs are considered to be normalized in Eq. (\ref{DynEq}) to be in the range of [0,1]. The obtained values for these inputs using the proposed control strategy are shown in Fig. \ref{Control inputs}, which satisfy the physiological constraint ($u_1 \in [0,1]$).  This implies that the considered desired scenarios (\ref{Desired1}) and (\ref{Desired2}) for reduction of unaware susceptible and chronically infected compartments comply with the control input limitations. Figure \ref{Par} illustrates the tuning of estimated parameters ($\hat{\theta}_1$ and $\hat{\theta}_2$) based on the designed adaptation laws (\ref{AdpLaw1}) and (\ref{AdpLaw2}) in the presence of 20\% uncertainty.\\
 
 \subsection{System response to different uncertainty levels}
\begin{figure}
\begin{center}
	{\includegraphics[scale=0.75]{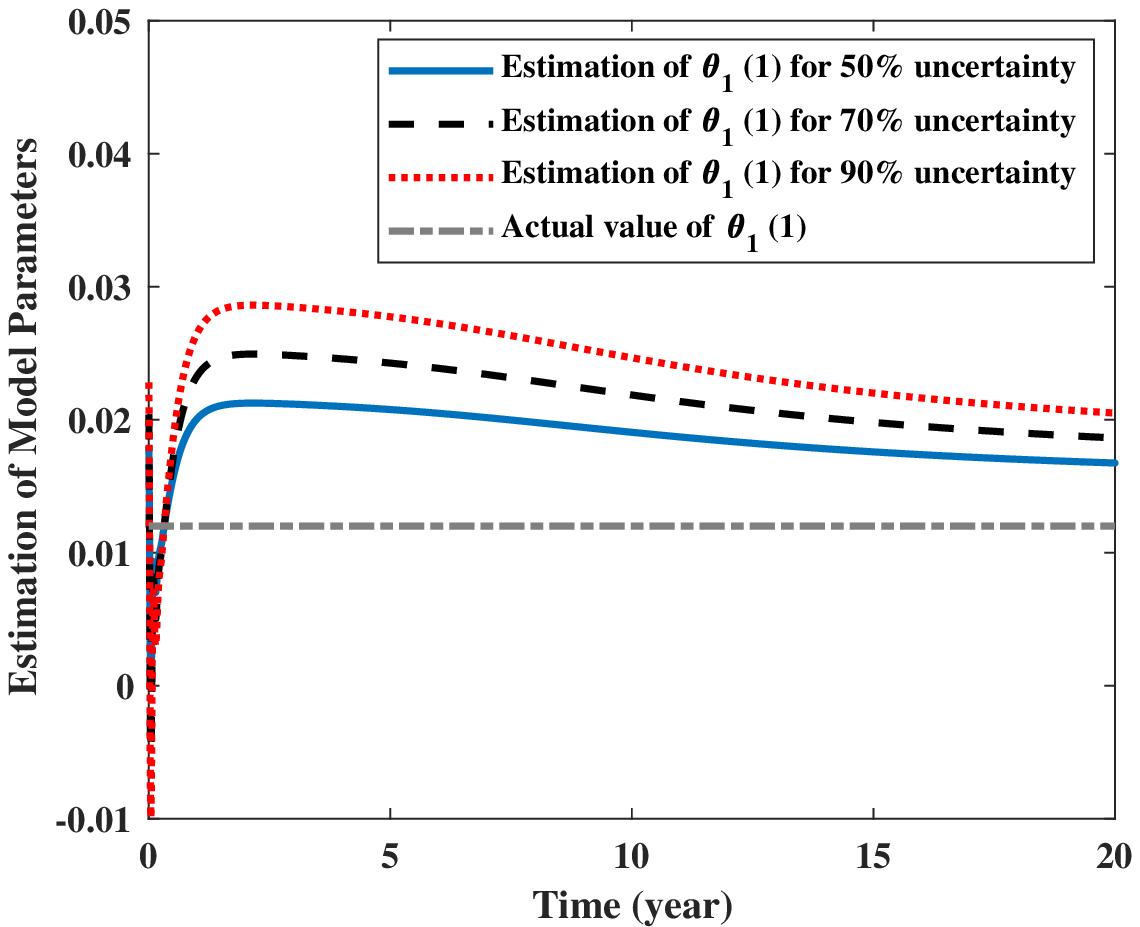}}\label{Uncer1}\\
	(a)\\
    {\includegraphics[scale=0.75]{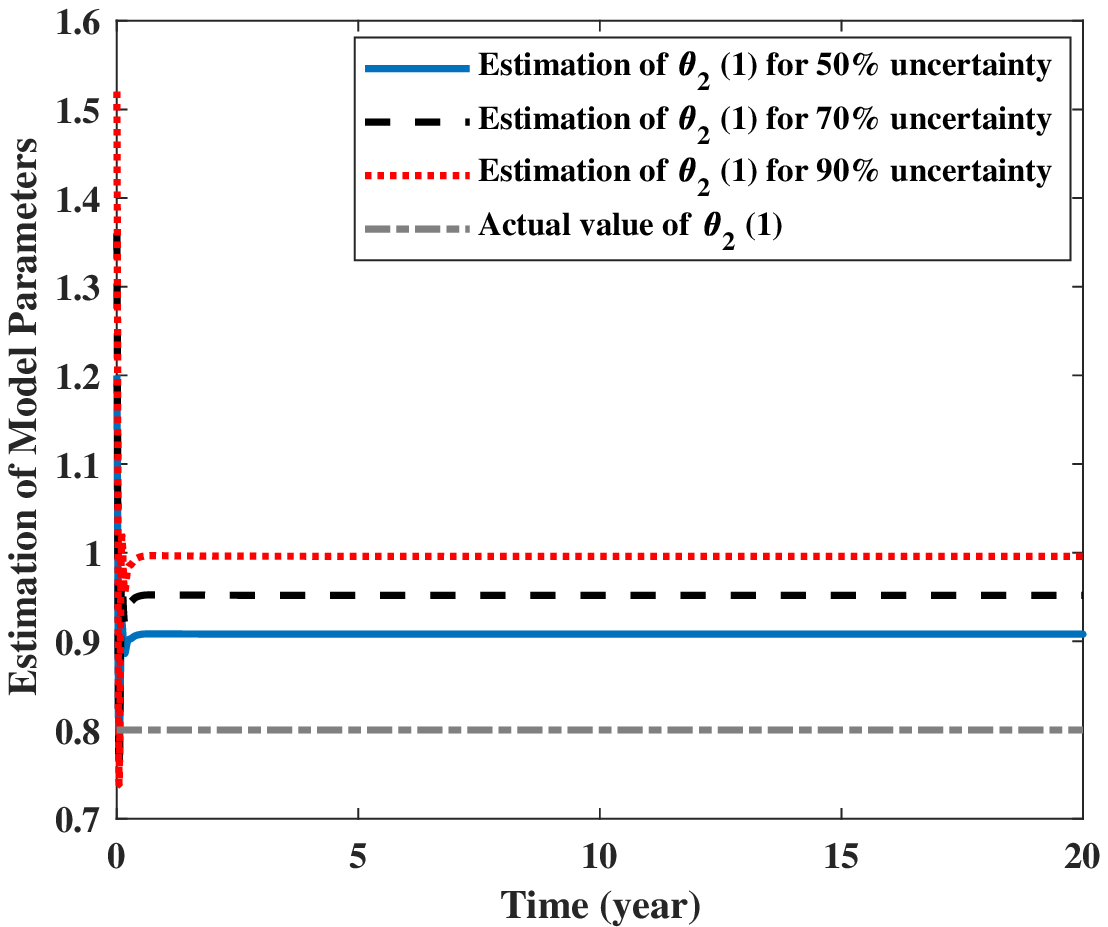}}\label{Uncer2}\\
    (b)

\caption{Adaptation of (a) $\theta_1(1)$ and (b) $\theta_2(1)$ using Eqs. (\ref{AdpLaw1}) and (\ref{AdpLaw2}), respectively, for different uncertainty levels} \label{Uncer}
\end{center}
\end{figure}

 In this section, effects of different uncertainty levels are investigated for the HCV epidemic dynamics. For this purpose, 50\%, 70\% and 90\% uncertainties are considered on the initial guess of parameters in $ \theta_1 $ and $ \theta_ 2$ (defined in Eqs. (\ref{RegPar1}) and (\ref{RegPar2})). Performance of the adaptation laws (\ref{AdpLaw1}) and (\ref{AdpLaw2}) on tuning of estimated model parameters is investigated in Fig. \ref{Uncer}. As discussed and  proven in Sec. \ref{sec3}, these adaptation laws guarantee that the estimation errors of the HCV dynamic parameters remain  bounded against different uncertainty levels. \\
 
\begin{figure}
 \begin{center}
 	{\includegraphics[scale=0.75]{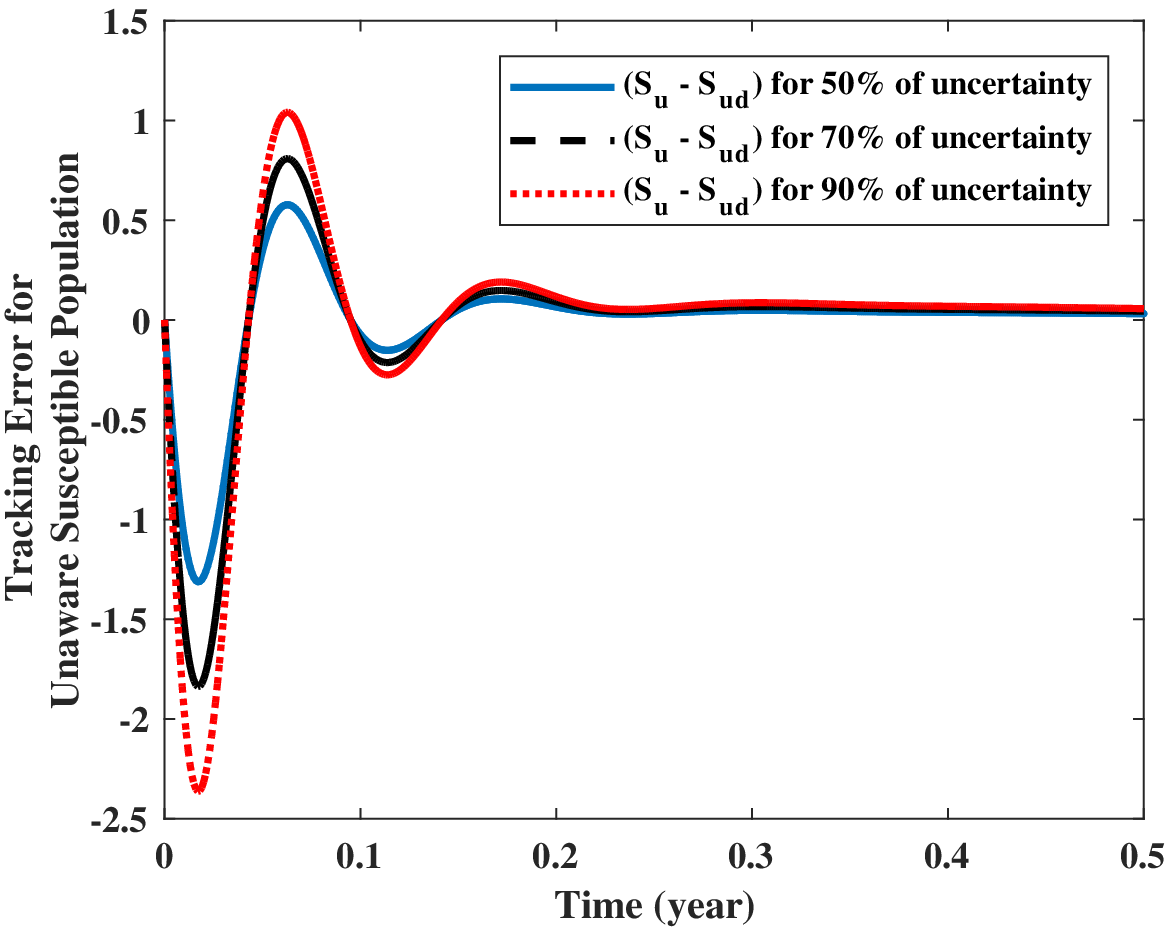}}\label{Uncerror1}\\
 	(a)\\
    {\includegraphics[scale=0.75]{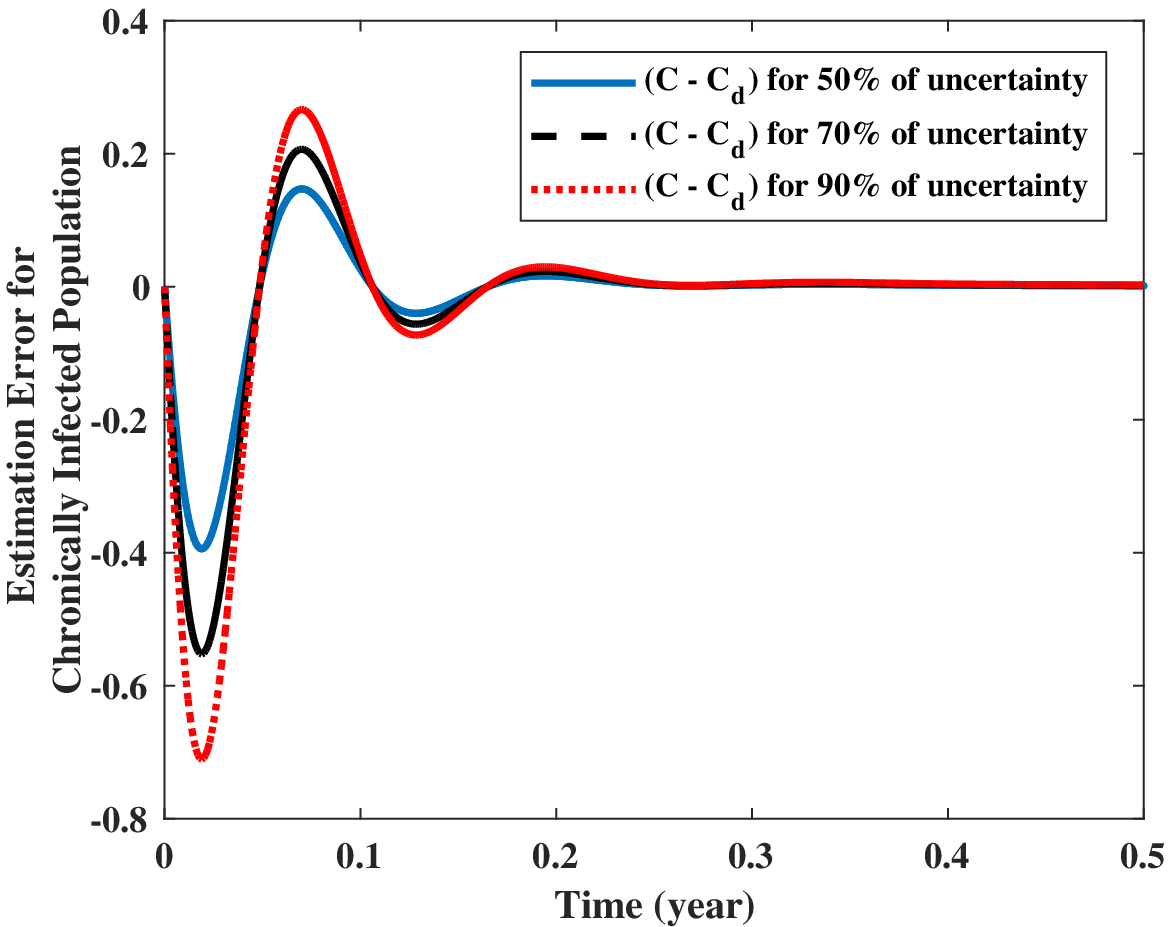}}\label{Uncerror2}\\
    (b)

 \caption{The difference between (a) unaware susceptible population and its desired value ($\tilde{S}_u=S_u - S_{u_d}$) and (b) chronically infected population and its desired value ($\tilde{C}=C-C_d$) for different parametric uncertainty levels} \label{Uncerror}
 \end{center}
 \end{figure}

 Figure \ref{Uncerror} shows the population errors of unaware susceptible and chronically infected classes in tracking their desired value ($\tilde{S}_u=S_u-S_{u_d}$ and $\tilde{C}=C-C_d$).\\
 
As observed in Fig. \ref{Uncerror} the increment of parametric uncertainties, increases the magnitude of errors ($\tilde{S_u}$ and $\tilde{C}$) and their initial variations. However, after a period of time (about 0.2 year), the errors magnitudes has reached zero, which means that the tracking convergence has been achieved for different values of uncertainties. In other words, the population of unaware susceptible and chronically infected compartments converged to their desired values ($S_u\rightarrow S_{u_{d}}$ and $C\rightarrow C_{d}$) in the existence of different levels of uncertainty.\\

\section{Conclusion}
In the present study, a new nonlinear adaptive strategy was developed to control the hepatitis C virus epidemic based on a mathematical model having uncertainties. For the first time, an adaptive feedback controller was employed to decrease the populations of unaware susceptible and chronically infected compartments based on the desired scenarios. Two control inputs were employed for this goal. The first one $u_1(t)$ is the effort rate to inform the susceptible individuals from the HCV and the second one $u_2(t)$ is the rate of treatment for chronically infected people. The Lyapunov stability theorem and the Barbalat's lemma were used to prove the tracking convergence to desired treatment scenarios. The proposed control laws and adaptation laws provided the stability of the closed-loop HCV epidemic system in the presence of parametric uncertainties. Results of numerical simulations showed that by adjusting the control inputs and the estimated parameters based on this strategy, number of the unaware susceptible and chronically infected individuals are decreased. As a result, population of the aware susceptible was increased and population of the acutely infected and treated classes reached out to zero at the end of process. Moreover, the obtained results implied that the tracking convergence is achieved for a wide range of uncertainties. Designing optimal trajectories and employing unstructured uncertainties can be considered as the next steps of this research in the future. \\

\appendix
\section{Barbalat's lemma}
\label{appendix-sec1}
The Lyapunov function \textit{V}($\tilde{S}_u$, $\tilde{C}$, $\tilde{\theta}_1$, $\tilde{\theta}_2$) in (\ref{Lyapunov}) is positive definite and its time derivative (\textit{$\dot{V}$}($\tilde{S}_u$, $\tilde{C}$)) in (\ref{Vdot2}) is negative semi definite. Thus, based on the Lyapunov stability theorem [\citenum{Slotine}], \textit{V} is bounded and it is concluded that $\tilde{S}_u$, $\tilde{C}$, $\tilde{\theta}_1$ and $\tilde{\theta}_2$ remain bounded.  \\

\textbf{Barbalat's lemma:} if g is a uniformly continues function and $\lim_{t\to\infty}{\int_{0}^{t}{g(\eta)d \eta}}$ exists and has a finite value, it is guaranteed that [\citenum{Slotine}]:\\
\begin{align}
\lim_{t\to\infty}{g(t)}=0 
\end{align}

In order to use this lemma for the HCV controlled system,   \textit{g(t)} is considered to be -\textit{$\dot{V}$}:\\
\begin{align}
g(t)=-\dot{V}=\lambda_1 \tilde{S}_u^2 + \lambda_2 \tilde{C}^2  \label{gVdot}
\end{align}

By integrating both sides of (\ref{gVdot}), one can write:\\
\begin{align}
V(0)-V(\infty)=\lim_{t\to\infty}{\int_{0}^{t}{g(\eta)d\eta}}
\label{V0-Vinf}
\end{align}

Since $\dot{V}$ is negative, \textit{$V(0)$} is larger than \textit{$V(\infty)$} and $V(0)-V(\infty)\geq0$. Moreover, as mentioned previously, \textit{V} is bounded based on the Lyapunov stability theorem . Thus, $\lim_{t\to\infty}{\int_{0}^{t}{g(\eta)d\eta}}$ in (\ref{V0-Vinf}) exists and has a bounded value. Therefore, it concluded using the Barbalat's lemma that:\\
\begin{align}
\lim_{t\to\infty}{(\lambda_1 \tilde{S}_u^2 + \lambda_2 \tilde{C}^2)}=0
\label{lim=0}
\end{align}

\bibliography{mybibfile}

\end{document}